\documentclass[apj]{emulateapj}
\usepackage{amsmath}
\usepackage{apjfonts}
\newcommand{\DDir}{\relax{D\kern-.7em{/}}}

\newcommand{\inv}[1]{\frac{1}{#1}}

\newcommand{\X}{\times}

\newcommand{\be}{\begin{equation}}
\newcommand{\ee}{\end{equation}}
\newcommand{\bea}{\begin{equation*}}
\newcommand{\eea}{\end{equation*}}

\newcommand{\nin}{\relax{\in\kern-.8em{/}}}

\newcommand{\al}{\alpha}
\newcommand{\bt}{\beta}

\newcommand{\lm}{\lambda}
\newcommand{\Lm}{\Lambda}
\newcommand{\de}{\delta}

\newcommand{\sig}{\sigma}

\newcommand{\vep}{\varepsilon}

\newcommand{\vs}{\textrm{v}_s}
\newcommand{\vt}{\textrm{v}}

\newcommand{\cm}{\mbox{ cm}}

\newcommand{\erg}{\mbox{ erg}}

\newcommand{\eV}{\mbox{eV}}
\newcommand{\keV}{\mbox{ keV}}
\newcommand{\MeV}{\mbox{ MeV}}

\newcommand{\G}{\text{G}}

\newcommand{\gr}{\mbox{g}}

\newcommand{\sref}{\S~\ref}

\newcommand{\TskkeV}{\left(\frac{T_s}{10\keV}\right)}
\begin{document}
\title{Fast radiation mediated shocks and supernova shock breakouts}
\author{Boaz Katz\altaffilmark{1}, Ran Budnik\altaffilmark{1}, and Eli
Waxman\altaffilmark{1}}

\altaffiltext{1}{Physics Faculty, Weizmann Institute, Rehovot 76100, Israel; ranny.budnik@weizmann.ac.il; boazka@wizemail.weizmann.ac.il, waxman@wicc.weizmann.ac.il}

\begin{abstract}
We present a simple analytic model for the structure of non-relativistic and relativistic radiation mediated shocks. At
shock velocities $\bt_s\equiv\vt_s/c\gtrsim 0.1$ the shock transition region is far from thermal equilibrium, since the
transition crossing time is too short for the production of a black-body photon density (by Bremsstrahlung emission). In
this region, electrons and photons (and positrons) are in Compton (pair) equilibrium at temperatures $T_s$ significantly
exceeding the far downstream temperature, $T_s\gg T_d\approx 2(\vep n_u \hbar^3c^3)^{1/4}$. $T_s\gtrsim 10\keV$ is
reached at shock velocities $\bt_s\approx 0.2$. At higher velocities, $\bt_s\gtrsim0.6$, the plasma is dominated in the
transition region by e$^\pm$ pairs and $60\keV\lesssim T_s \lesssim 200\keV$. We argue that the spectrum emitted during
the breaking out of supernova shocks from the stellar envelopes (or the surrounding winds) of Blue Super Giants and
Wolf-Rayet stars, which reach $\bt_s>0.1$ for reasonable stellar parameters, may include a hard component with photon
energies reaching tens or even hundreds of $\keV$. Our breakout analysis is restricted to temperatures $T_s\lesssim 50\keV$ corresponding to photon energies $h\nu\lesssim 150\keV$, where pair creation can be neglected. This may account for the X-ray outburst associated with SN2008D, and
possibly for other SN-associated outbursts with spectra not extending beyond few $100\keV$ (e.g. XRF060218/SN2006aj).
\end{abstract}
\keywords{~~~shock waves --- radiation mechanisms: nonthermal --- X-rays: bursts --- supernovae: general : individual (SN 2008D)}

\section{Introduction}
During core collapse supernova explosions, strong shock waves traverse the exploding stars' mantle/envelope. These shocks move with high velocities, $\bt_s\equiv\vt_s/c\gtrsim 0.01$, through dense, $n\sim 10^{15}$ to $10^{20}\cm^{-3}$, optically thick plasma.
Under such conditions, the shocked plasma's thermal energy density is dominated by radiation as long as
\begin{equation}
\bt_s\gg\left(\frac{n}{a_{BB}}\right)^{1/6}(m_pc^2)^{-1/2}\sim 3\times 10^{-4} \left(\frac{n}{10^{20}\cm^{-3}}\right)^{1/6},
\end{equation}
where $a_{BB}=\pi^2/15(\hbar c)^{-3}$ is the Stefan-Boltzman energy density coefficient.
In such shocks, the photons that are generated in the downstream region diffuse upstream and decelerate the incoming plasma by colliding with the electrons, which in turn stop the nuclei through collective plasma interactions (e.g. an electrostatic field).
The typical width of such Radiation Mediated Shocks (RMS) is the length photons can diffuse into the upstream $l_d\sim 1/n_u\bt_s\sig_T$, where $\sig_T$ is the Thompson cross section, and the subscript $u$ denotes in this paper the upstream values of parameters.

The structure of non-relativistic radiation mediated shocks was studied by \citet{Colgate74} and \citet{Weaver76},
motivated by the idea that the post shock plasma can reach high temperatures, $T\gtrsim 10\MeV$, leading to the
generation of deuterium, as an alternative to Big Bang nucleosynthesis. Non relativistic numerical calculations were
preformed by \citet{Weaver76} for shock velocities  $\bt_s<0.3$. Self-Consistent solutions of the hydrodynamic, radiation
generation (dominated by Bremsstrahlung) and diffusion equations were obtained, confirming that RMS are consistent
physical structures. At low velocities, $\bt_s\lesssim 0.1 c$ [c.f. equation \eqref{eq:btNeq}] thermal equilibrium is
maintained throughout the shock. In such conditions the shock temperature monotonically increases between the far
upstream and the far downstream values \citep[e.g.][]{Zel'dovich66}. For higher velocities, it was found that the plasma
in the shock transition region departs from thermal equilibrium, and the electron temperature reaches values of tens of
$\keV$, greatly exceeding the far downstream values (but much smaller than the suggested $10\MeV$ temperatures). The
results of the non-relativistic calculations are not applicable to shocks velocities exceeding $\beta_s\simeq0.2$, where
the high temperatures reached in the shock transition require inclusion of relativistic processes (e.g. pair production),
and where the evolution of the photon distribution can no longer be simply described as diffusion in space and momentum.

Detailed studies of Compton scattering in the converging flow of non-relativistic RMS [\citet{Blandford81b}, \citet{Lyubarskii82}, \citet{Becker88} and recently in the context of shock-breakouts, \citet{Wang07}] have shown that photons traversing through RMS can significantly increase their energy while scattering back and forth in the converging flow in a manner similar to Fermi-acceleration of Cosmic rays.
The energy photons can reach by such 'bulk' acceleration is limited to
\begin{equation}\label{eq:numaxBP}
h\nu_{\max}\sim 0.2m_ec^2\bt_s^2\sim 1(\bt_s/0.1)^2\keV,
\end{equation}
above which the energy loss per collision due to the Compton recoil is larger than the energy gain due to the converging bulk flow. As we show here, for velocities $\bt_s\gtrsim 0.1$ the
electrons (and positrons) are "heated" in the immediate downstream (the region where the photons are generated) to energies exceeding those given in Eq. \eqref{eq:numaxBP} [c.f. Eq.\eqref{eq:btOfT}]. Thus, photons are generated with typical energies larger than  given in Eq. \eqref{eq:numaxBP}, implying that 'bulk' photon acceleration is only relevant for lower shock velocities, $\bt_s\lesssim 0.1$, and for sub-$\keV$ photons (and for densities low enough so that the downstream thermal temperatures are considerably smaller than $h\nu_{\max}$).

In this paper we present a simple analytic model for the structure of radiation mediated shocks. The model accurately
reproduces the numerical results of \citet{Weaver76} for $\beta_s\lesssim0.2$, and provides an approximate description of
the shock structure at larger velocities, $\beta_s\rightarrow1$. A detailed comparison of the results of the simple model
presented here with the results of exact solutions of the shock structure (self-consistently solving the photon transport
equation and the plasma flow equations for mildly and highly relativistic shocks)  will be presented in a followup paper
\citep{Budnik10}. We assume in our calculations that photons are generated mainly by Bremsstrahlung emission. This
assumption is valid in the absence of strong magnetic fields (note that direct bound-free transitions do not lead to net
generation of photons while the contribution of bound-bound transitions is negligible under the conditions relevant to
shock-breakout).

We first present in \sref{sec:NRRMS} an analytical model for Non-Relativistic RMS, deriving simple analytic approximations for the non-equilibrium temperature and for the width of the temperature profile, and then extend our model in \sref{sec:RRMS} to relativistic RMS. In section \sref{sec:Breakouts} we discuss the implications of our results to the expected properties of SN X-ray outbursts. We show that breakout velocities $\bt_s>0.1$ may be reached  in the explosions of Blue Super Giant (BSG) and Wolf-Rayet (WR) stars (\sref{sec:BreakoutDynamics}), and argue that the X-ray outbursts accompanying breakouts from such stars may include a hard component with photon energies reaching tens or even hundreds of $\keV$ (\S~\ref{sec:BreakoutEmission}). The implications of our results to the interpretation of recent X-ray outbursts associated with SNe (XRO080109/SN2008D, XRF060218/SN2006aj) are discussed in \S~\ref{sec:observations}.
We summarize the results and discuss their implications
in \sref{sec:Discussion}.

\section{Non-Relativistic radiation mediated shocks}\label{sec:NRRMS}
Consider a radiation mediated shock in a plasma consisting of protons, electrons and photons.
The far upstream consists of cold protons and electrons moving with velocity $\bt_s c$ in the shock rest frame, while the far downstream consists of protons, electrons and photons in thermal equilibrium with a temperature $T_d$.

In this section we construct an analytical, approximate model for the steady state profile of such a shock. We start by writing down the downstream conditions in \sref{sec:DonstreamCond}. Next, we discuss the velocity profile in \sref{sec:Deceleration} and argue that the deceleration scale is roughly $(\bt_sn_u\sig_T)^{-1}$. In \sref{sec:LT} we show that for large shock velocities, $\bt_s\gtrsim 0.1$, the temperature can have a non trivial profile extending to distances much larger than the deceleration length.
In \sref{sec:ImmediateDownstream} we estimate the temperature in the shock transition and show that it can greatly exceed $T_d$. We show that our simple analytic results are in good agreement with detailed numerical calculations. Finally, we use the results derived in \S~\ref{sec:DonstreamCond}--\S~\ref{sec:ImmediateDownstream} to describe in \sref{sec:ShockStructure} the structure of fast NR radiation mediated shocks.

\subsection{Downstream conditions}\label{sec:DonstreamCond}
The energy and momentum flux are dominated by the protons' momentum
and kinetic energy in the upstream and by the photons in the far
downstream. Proton number, momentum and energy conservation require
that
\begin{align}\label{eq:udconservation}
&n_u\bt_s=n_d\bt_d,\cr
&p_{\gamma,d}=n_u\bt_s(\bt_s-\bt_d)m_pc^2,\cr
&4p_{\gamma,d}\bt_d=n_u\bt_s(\bt_s^2-\bt_d^2)m_pc^2/2,
\end{align}
where the subscript $d$ denotes the downstream values of the parameters and we used the fact that the energy density in photons is $e_{\gamma}=3p_\gamma$.
Equations \eqref{eq:udconservation} imply $\bt_d=\bt_s/7$. Thermal equilibrium implies
\begin{equation}
e_{\gamma,d}=a_{BB}T_d^4.
\end{equation}
The far downstream temperature is thus
\begin{align}\label{eq:Td}
&T_d=\left(\frac{315}{4\pi^2}n_u\vep\hbar^3c^3\right)^{1/4}\approx 0.16 (\vep_{1}n_{u,15})^{1/4} \keV,\cr &\approx 0.13
n_{u,15}^{1/4}\bt_{s,-1}^{1/2}\keV\cr
\end{align}
where $n_u=10^{15}n_{u,15}$, $\vep=0.5\bt_s^2 m_p c^2=10\vep_1\MeV$ and $\bt_s=0.1\bt_{s,-1}$.

\subsection{Deceleration}\label{sec:Deceleration}
Next we estimate the proton deceleration length scale. Once a proton reaches a point in the shock where the energy is dominated by photons, it experiences an effective force
\begin{equation}\label{eq:protondrag}
 \bt_s\frac{d\bt}{dx}m_p c^2\sim \sig_T\bt_s e_\gamma \sim \sig_T n_u\bt_s^3m_pc^2,
\end{equation}
implying a deceleration length of
\begin{equation}\label{eq:ShockWidth}
L_{\text{dec}}\equiv\bt_s \left(\frac{d\bt}{dx}\right)^{-1}\sim \inv{\sig_Tn_u\bt_s}.
\end{equation}
We note that once the photons dominate the pressure, they
cannot drift with the protons, as this will imply an energy flux
greater than the total energy flux. Thus, the drag estimated in
equation \eqref{eq:protondrag} is unavoidable. The energy and
momentum of the protons are thus transferred to the photons on a
length scale $L_{\text{dec}}$ irrespective of the details of the
mechanisms that generate the energy in the radiation field. We note
that the transition length, $L_{\text{dec}}$, is of the order of the
distance a photon can diffuse upstream before being advected with
the flow. The deceleration of the flow can be solved analytically
\citep[e.g.][and references within, cf Eq. \eqref{eq:AanalyticSolution}]{Blandford81b}. For the strong
shocks considered here,
\begin{equation}\label{eq:analyticSolution}
x=\frac{1}{21\sig_Tn_u\bt_s}\ln\left[\frac{(\bt_s-\bt)^7}{(7\bt-\bt_s)\bt_s^6}\right],
\end{equation}
where $x$ is the distance measured in the shock frame and the point $x=0$ corresponds to $\bt/\bt_d-1\approx 0.25$.

\subsection{Temperature profile length scale}\label{sec:LT}
The region of the shock profile over which the temperature changes before it reaches $T_d$ can be extended to distances that are much larger than $L_{\text{dec}}$.

To see this, consider the length scale that is required to generate the density of photons of energy $\sim T_d$ in the downstream, determined by thermal equilibrium, $n_{\gamma,\text{eq}}\approx p_{\gamma,d}/T_d$,
\begin{equation}\label{eq:LTdefinition}
L_{T}\sim \bt c\frac{n_{\gamma,\text{eq}}}{Q_{\gamma,\rm eff.}},
\end{equation} where $Q_{\gamma,\rm eff.}$ is the \textit{effective} generation rate of photons of energy $3T_d$. We use here the term "\textit{effective} generation rate" due to the following important point. Photons that are produced by the emission mechanism at energies $\ll T_d$ may still be counted as contributing to the production of photons at $T_d$, since they may be upscattered by inverse-Compton collisions with the hot electrons to energy $\sim T_d$ on a time scale shorter than that of the passage of the flow through the thermalization length, $L_T/\beta_d c$. Note, that the relevant Compton $y$ parameter,
\begin{align}
&y=\frac{4T}{m_ec^2}L_Tn_d\sig_T\bt_d^{-1}=49\frac{4T}{m_ec^2}(L_T\bt_sn_u\sig_T)\bt_s^{-2},\cr &\sim 4
(L_T\bt_sn_u\sig_T)\frac{T}{100 \eV}\bt_{s,-1}^{-2},
\end{align}
is much larger than one for $L_T\bt_sn_u\sig_T\gg 1/4 (T/100 eV)^{-1}\bt_{s,-1}^{2}$. $Q_{\gamma,\rm eff.}$ includes
therefore all photons produced down to an energy that allows them to be upscattered to $T_d$. For Bremsstrahlung
emission, which we assume to be the main source of photons, the number of photons generated diverges logarithmically at
low energy, so that $Q_{\gamma,\rm eff.}$ may be significantly larger than the Bremsstrahlung generation rate of photons
at $T_d$.

In order for the photon energy to be significantly increased by scattering, it must be scattered $\sim m_ec^2/(4T)$ times before getting re-absorbed. This sets a lower limit to the frequency of photons that should be included in $Q_{\gamma,\rm eff.}$,
\begin{align}\label{eq:AbsorptionCutoff}
h\nu_{a}&=m_ec^2\left(\frac{T}{m_e c^2}\right)^{-5/4}\left(\frac{\al_e g_{\text{ff}} (m_ec^2)^{-3}h^3 c^3 n}{32\pi}\right)^{1/2}\cr
&\sim 60g_{\text{ff}}^{1/2}\left(\frac{T}{100\eV}\right)^{-5/4}n_{15}^{1/2}\eV,
\end{align}
where $n=7n_u=7\times 10^{15}n_{15}\cm^{-3}$.
The generation rate should include all the photons that are generated with energy above $h\nu_a$ and are able to be upscattered during the available time $L_{T}/(\bt c)$.
The Bremsstrahlung effective photon generation rate is thus given by
\begin{equation}\label{eq:QgammaBr}
Q_{\gamma,\text{eff}}=\al_e n_pn_e\sig_T c\sqrt{\frac{m_ec^2}{T}}\Lm_{\text{eff}}g_{\text{eff}},
\end{equation}
where $g_{\text{eff}}$ is the Gaunt factor and $\Lm_{\text{eff}}\sim\log[T/(h\nu_a)]$. As long as the Compton $y$
parameter is large $y>\log(T/h\nu_a)$, all photons produced above $\nu_a$ are upscattered to $\sim 3T$ energies.

Using equations \eqref{eq:QgammaBr} and \eqref{eq:LTdefinition} we obtain
\begin{align}\label{eq:LT}
L_T\bt_sn_u\sig_T&\sim \inv{100\al_e\Lm_{\text{eff}}g_{\text{eff}}}\frac{\vep^2}{\sqrt{m_e c^2 T}m_pc^2}\cr
&\approx 16 \vep_{1}^{15/8}n_{15}^{-1/8}\Lm_{\text{eff}}^{-1}g_{\text{eff}}^{-1}.
\end{align}
This implies that for large shock velocities,
\begin{equation}\label{eq:btNeq}
\bt_s>0.07 n_{15}^{1/30}(\Lm_{\text{eff}}g_{\text{eff}})^{4/15},
\end{equation}
the length required to produce the downstream photon density is much larger than the deceleration scale. For lower shock velocities, thermal equilibrium is approximately maintained throughout the shock.

\subsection{Immediate downstream}\label{sec:ImmediateDownstream}
The photons that stop the plasma at the deceleration region must be generated within a distance
$\sim(n_u\bt_s\sig_T)^{-1}$ of the deceleration region in order to be able to reach it (before being swept downstream by
the flow). Consider the following simple model for the first few diffusion lengths downstream of the deceleration region.
We assume that the velocity is $\bt_d$ downstream from some point $x=x_0$ and is much larger upstream of it. As the
e-folding distance of the residual velocity $\delta\bt=\bt-\bt_d$ is very short, $=\bt_sn_u\sig_T/21$, this is a good a
approximation. We neglect any generation of photons upstream of the transition, $Q_{\gamma,\rm eff}(x<x_0)=0$. This is
reasonable as the contribution of photons from a slab, one diffusion length wide, is roughly proportional to $\bt^{-3}$
and $\bt$ changes by a factor of 7 along the transition region. We further assume that the generation rate of photons
downstream of the transition is roughly constant and equal to $Q_{\gamma,\rm eff}(x=x_0)$. The latter simplification is
justified, to an accuracy of a few tens of percents, due to the fact that the temperature changes by order unity across
the available (diffusion length) distance and $Q_{eff}\propto T^{0.5}$. For such a flow, the density of effective photons
at $x=x_0$ is:
\begin{equation}\label{eq:ngamma0}
n_\gamma(x_0)=\inv{\bt_d^2 c^2}Q_{\gamma,\rm eff}\frac{c}{3 n_d \sig_T}.
\end{equation}
As the flow at this point is close to the downstream velocity, we
have $n_\gamma(x_0)T_s=p_{\gamma,d}$ where $T_s\equiv T(x_0)$ is the
temperature in the immediate downstream. Using Eqs
\eqref{eq:QgammaBr} and \eqref{eq:ngamma0}, the momentum
conservation equation can be written as
\begin{equation}\label{eq:MainEquation}
\inv{3}\al_e\Lm_{\text{eff}}g_{\text{eff}}\sqrt{m_e c^2 T_s}n_d\bt_d^{-2}=\frac67n_u\bt_s^2m_pc^2,
\end{equation}
where $\Lm_{\text{eff}}=10\Lm_{\text{eff},1}\sim\log[T_s/(h\nu_a)]$
with $h\nu_a$ given by Eq.~\eqref{eq:AbsorptionCutoff}. Here we
assumed that the temperature does not change much within a distance
of $L_\mathrm{dec}$ downstream of the point $x=x_0$. The Compton $y$
parameter, relevant for the upscattering of the photons that diffuse up to the
deceleration region,
\begin{align}
&y=\frac{4T_s}{m_ec^2}\bt_d^{-2}\sim 400 \TskkeV \bt_{s,-1}^{-2},\cr
\end{align}
is much larger than 1 for $T_s\gg 30 \bt_{s,-1}^{2}\eV$. We therefore assume that all photons above $h\nu_a$ are
upscattered all the way to $T_s$.

Solving Eq. \eqref{eq:MainEquation} for $\bt_s$ we find:
\begin{align}\label{eq:btOfT}
\bt_s&=\frac{7}{\sqrt{3}}\left(\inv{2}\al_e \Lm_{\text{eff}}g_{\text{eff}}\right)^{1/4}\left(\frac{m_e}{m_p}\right)^{1/4}\left(\frac{T_s}{m_ec^2}\right)^{1/8}\cr
&\approx 0.2\Lm_{\text{eff},1}^{1/4}\left(\frac{g_{\text{eff}}}{2}\right)^{1/4}\TskkeV^{1/8}.\cr
\end{align}
Eq. \eqref{eq:btOfT} is in good agreement with the results of \citep{Weaver76} for $T_s<50\keV$. For $\bt_s>0.07$ and $n_u=10^{15}\cm^{-3}-10^{22}\cm^{-3}$ the temperature (velocity) given by equation~(\ref{eq:btOfT}) for a given velocity (temperature) agrees with the numerical calculations to within a factor of $1.5$ (20\%) \citep[with the gaunt factor calculated as in][]{Weaver76}. We note that the scaling with parameters in equation \eqref{eq:btOfT} can be found by writing the RMS equations in dimensionless form (cf Eqs. \eqref{eq:ATnorm}-\eqref{eq:AtildeTn}).

Comparing Eq. \eqref{eq:btOfT} with the equation for the far downstream temperature $T_d$, eq.~(\ref{eq:Td}), we see that the shock temperature is much larger than the downstream temperature for shock temperatures $T_s\gtrsim 1\keV$ corresponding to shock velocities $\bt_s\gtrsim 0.1$ \citep[see also][]{Weaver76}.

Note, that for temperatures $T\gtrsim 1\keV$ the typical photons in the immediate downstream have energies $h\nu\sim 3T$ higher than the maximal energies attainable by bulk Compton acceleration in non-relativistic flows, given by Eq.
\eqref{eq:numaxBP}. Indeed, the typical photon energies, as derived from \eqref{eq:btOfT}, rise much faster with velocity than $h\nu_{\max}\propto \bt_s^2$, and at $T\sim 10\keV$, corresponding to $\bt_s\approx 0.2$, are already much higher ($h\nu\sim 3T\sim 30\keV$ compared to $h\nu_{\max}\sim 4\keV$). At temperatures exceeding $T\sim 50\keV$, where pair production becomes important, Eq. \eqref{eq:MainEquation} is not applicable anymore. However, the velocity required for bulk Comptonization to be able to accelerate to the typical photon energies at those temperatures, $h\nu\sim 150 \keV$, approaches $c$. Thus, for non-relativistic and mildly relativistic shock velocities with $\bt_s\gtrsim 0.1$, bulk photon acceleration cannot reach energies that are considerably higher than the typical "thermally"-Comptonized photon energies in the immediate downstream.

\subsection{Description of the shock structure}\label{sec:ShockStructure}
We can broadly divide the shock into four separate regions.
\begin{enumerate}
\item Near upstream: A few diffusion lengths, $(\bt_s\sig_T n_u)^{-1}$, upstream of the deceleration region. In this region, characterized by velocities that are close to the upstream velocity, $\bt\approx \bt_s$, and temperatures  $T\gg T_u$, the temperature changes from $T_u$ to $\sim T_{s}$. It ends when the fractional velocity decrease becomes significant.
\item Deceleration region: A $(\bt_s\sig_T n_u)^{-1}$ wide region where the velocity changes from $\bt_{u}$ to $\bt_{d}$ and the temperature is roughly constant, $T\simeq T_s$.
\item Immediate downstream:  Roughly a diffusion length, $(\bt_s\sig_T n_u)^{-1}$, downstream of the deceleration region. In this region, characterized by velocities close to the downstream velocity, $\bt\approx\bt_d$, and temperature $T\sim T_s$, the photons that stop the incoming plasma are generated. Upstream of this region $\bt>\bt_d$ and the photon generation rate is negligible. Photons that are generated downstream of this region are not able to propagate up to the transition region.
\item Intermediate downstream: The region in the downstream where most of the far downstream photons are generated and $T$ changes from $T_s$ to $T_d$. This region has a width $L_T$ given by Eq. \eqref{eq:LT}, much greater than $(\bt_s\sig_T n_u)^{-1}$. Thus diffusion within this region can be neglected. The temperature profile is expected to follow $T\propto x^{-2}$. To see this, note that the photon density at a distance $x$ from the shock is proportional to the integral of the photon generation, $n_\gamma\propto T^{-1/2}x$. Since the photon pressure equals the downstream pressure, we have $n_\gamma\propto T^{-1}$ and $T\propto x^{-2}$ (this is valid for a constant value of $\Lm_{\text{eff}}g_{\text{eff}}$ and is somewhat shallower in reality). Using this dependence of the temperature on distance, we find that $L_T\bt_sn_u\sig_T\sim \Lm_{\text{eff}}g_{\text{eff}}|_dT_d^{-1/2}[(\Lm_{\text{eff}}g_{\text{eff}})|_sT_s^{-1/2}]^{-1}$, in agreement with equations \eqref{eq:LT} and \eqref{eq:btOfT}.
\end{enumerate}

\begin{figure}[h]
\epsscale{1} \plotone{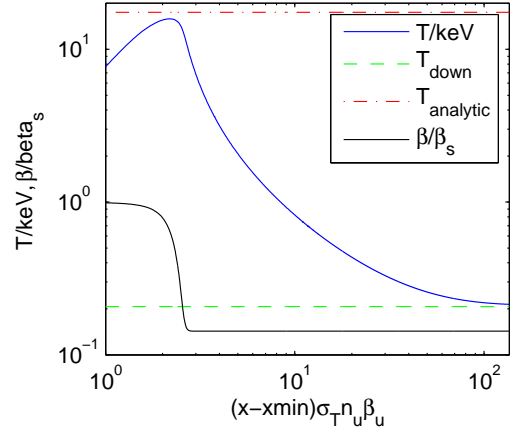}
\caption{\label{fig:Shock} RMS Shock temperature and velocity profile for $\vep=30\MeV$ ($\bt_s=0.25$) and $n_u=10^{15}\cm^{-3}$. The black full line is the velocity given by \eqref{eq:analyticSolution}. The Blue full line is the temperature calculated numerically [see \sref{sec:DetailedSol}]. The dashed green line is the far downstream equilibrium temperature given by Eq. \eqref{eq:Td}. The dashed-doted red line is the analytic estimate for the immediate downstream temperature. The width of the temperature equilibrium length, given by eq. \eqref{eq:LT}, for these parameters is $L_T\bt_u\sig_T n_u\approx 130$.}
\end{figure}

As an illustration, the profiles of the velocity and temperature of a non-relativistic RMS ($\bt_s=0.25,n_u=10^{15}\cm^{-3}$) are shown in figure \ref{fig:Shock}. A description of a novel method used to find such solutions is described in \sref{sec:DetailedSol}. As can be seen, the analytic estimates are in good agreement with the numerical calculation.

\section{Relativistic radiation mediated shocks}\label{sec:RRMS}

\subsection{General discussion}
The analysis presented above is valid for temperatures up to about $T_s\sim30\keV$, where a number of effects that were
not taken into account begin to affect the temperature profile. The main process that should be taken into consideration
is the creation of electron-positron pairs at $T\gtrsim 50\keV$. Additional corrections include relativistic corrections
to the Bremsstrahlung emission rate, inclusion of photon creation by double Compton scattering, Klein-Nishina (KN)
corrections to the Compton cross section, reduction of the Compton $y$ parameter due to the smaller optical depth,
$\tau\sim\bt_d^{-2}$, and the anisotropy due to the high velocities, $\bt_s\sim 1$.

The following important characteristic of the shock structure is likely unchanged: the plasma decelerates due to
scattering with photon and pairs that are produced in the first $\bt_d^{-1}$ optical depths of the downstream [a
discussion of the deceleration velocity profile, neglecting KN corrections, is given in \citet{Levinson08}, note that KN
corrections and significant electron-positron production significantly affect the result \citep{Budnik10}]. As the
downstream velocity is $\bt_d<1/3$ for all $\bt_s$, the diffusion approximation in the downstream is reasonable and an
equation similar to Eq. \eqref{eq:MainEquation}, with the relativistic corrections included, can be used to constrain the
temperature in the transition region.

Note that \citet{Weaver76} claimed that once a significant amount of pairs is created, the pairs will 'short out' the
electrostatic field which is responsible for the deceleration of the protons. He claimed that the electrons and ions will
no longer be fully coupled, and suggested that once there are many pairs per proton, there will be a relative drift of
order $\bt_s$ between the leptons and protons, and that the main mechanism for stopping the protons will be Coulomb
collisions with the leptons. We find this picture to be unrealistic in real plasmas. A relatively weak magnetic field,
which will be freezed into the pair plasma, will suffice to deflect the protons on scales which are much shorter than
determined by Coulomb collisions. Several length scales may be important here. The gyroradius of the protons in a
magnetic field $B=10^2B_2\G$ is
\begin{equation}
R_L=\gamma\bt\frac{m_pc^2}{eB}\sim 3\times 10^4 \gamma\bt B_2^{-1}\cm,
\end{equation}
the skin-depth of the pair plasma is
\begin{equation}
l_{sd}=\sqrt{\frac{m_ec^2}{4\pi e^2n_l}}\sim 1.7\X 10^{-2}n_{l,15}^{-1/2}\cm,
\end{equation}
and the Compton mean free path is,
\begin{equation}
l_{T}=\frac{1}{n_l\sig_T}\sim 1.5\times 10^9 n_{l,15}^{-1}\cm.
\end{equation}
As long as the Compton mean free path is much larger than the other two scales, it is probably safe to assume that the protons and electrons are highly coupled on the dynamical scales.
We thus assume in what follows that collective plasma effects will keep the pair plasma and the protons coupled without the need for an electrostatic field.

\subsection{T-$\bt_d$ relation}
As long as the Compton $y$ parameter is much larger than unity, the spectrum will be close to a Wein spectrum.
For the entire parameter space considered here, the positron number is in equilibrium. To see this, note that the typical time it takes a positron to annihilate is similar to the typical time it takes a photon to Compton scatter, thus pairs can be annihilated (or generated) in $\bt_d$ Compton optical depths compared with the $\bt_d^{-1}$ available.

A detailed discussion of the emission of plasmas in pair and Compton equilibrium is given by \citet{Svensson84}, who shows that the main emission processes are Bremsstrahlung and Double Compton, with a domination of Bremsstrahlung at temperatures above $T\approx 60\keV$.

A precise determination of the relation between $T_s$ and $\beta_u$ is complicated in the range $T_s\gtrsim 30\keV$ due to the relativistic corrections to the emission rates and due to the order unity $y$ parameter. Such determination requires a careful analysis of these effects. We avoid such a detailed analysis by limiting our analysis to a demonstration of the validity of the following statements, which set stringent constraints on the relation between $T_s$ and $\bt_d$:
\begin{enumerate}
\item Temperatures $T_s> 30\keV$ are reached for downstream velocities $\bt_d>0.03$, with $\bt_d\approx 0.03$ for $T_s\approx 30\keV$ (as shown in \S~\ref{sec:NRRMS});
\item At $\bt_d=0.1$ the shock transition temperature satisfies $60\keV<T_s$;
\item $T_s\lesssim 200\keV$ for all the possible downstream velocities, $\bt_d<1/3$;
\end{enumerate}
The validity of statements 2 and 3 is demonstrated in \S~\ref{sec:b0.1} and \S~\ref{sec:bs1} below by comparing the
photon to pair ratio that is determined by the balance of photon production and diffusion, eq.~\eqref{eq:ngamma0}, with
the ratio that is determined by pair production equilibrium. The complications due to the various corrections are avoided
since, under the conditions we consider, the following hold: The number of pairs greatly exceeds the number of protons,
$n_l\gg n_p$, where $n_l=n_++n_-\approx 2n_-$; Double Compton emission is negligible; The Compton $y$ parameter is large.

\subsection{At $\bt_d\approx0.1$, $60\keV<T_s$}
\label{sec:b0.1}

Assuming pair production domination and neglecting Double Compton emission, Eq.~\eqref{eq:ngamma0} can be written as
\begin{equation}\label{eq:MainEquation_rel}
\frac{n_\gamma}{n_l}=\inv{3}\al_e\Lm_{\text{eff}} \bar g_{\text{eff,rel}}(\hat T)\bt_d^{-2},
\end{equation}
where we wrote the free-free emission in the form
\begin{equation}\label{eq:QffRel}
Q_{\gamma,\text{eff}}=\al_e \sig_T c n_l^2 \Lm_{\text{eff}}\bar g_{\text{eff,rel}}(\hat T).
\end{equation}
Here, $\hat T\equiv T/(m_ec^2)$, and $\bar g_{\text{eff,rel}}$ is the total Gaunt factor [defined by Eq.
\eqref{eq:QffRel}] including all lepton-lepton Bremsstrahlung emission. For $10<\Lm_{\text{eff}}<20$ and
$60\keV<T<m_ec^2$ the approximation
\begin{equation}\label{eq:fBrApp}
\bar g_{\text{eff,rel}}\approx\Lm_{\text{eff}}/2,
\end{equation}
agrees with the results of \citet{Svensson84} to an accuracy of better than $25\%$. Substituting Eq.~\eqref{eq:fBrApp} in
Eq.~\eqref{eq:MainEquation_rel} we find
\begin{equation}\label{eq:PhotonGenerationRel}
\frac{n_\gamma}{n_l}\approx 10 \left(\frac{\Lm_{\text{eff}}}{10}\right)^2\bt_{d,-1}^{-2}.
\end{equation}
We next estimate the absorption frequency. The absorption is dominated by $e+e-$ Bremsstrahlung and for $T\gtrsim
0.1m_ec^2$ photons require a single scattering to change their energy considerably. The absorption frequency is thus
roughly equal to the threshold frequency $\nu_{C=B}$ at which the Compton scattering rate equals the absorption
scattering rate and is given by
\begin{align}\label{eq:AbsorptionCutoffRel}
h\nu_{a}\sim h\nu_{C=B}&=m_ec^2\left(\frac{T}{m_e c^2}\right)^{-3/4}\left(\frac{\al_e g_{\text{ff}} (m_ec^2)^{-3}h^3 c^3
n_l}{8\pi}\right)^{1/2}\cr &\sim 0.6 \left(\frac{g_{\text{ll,ff}}}{10}\right)^{1/2}\left(\frac{T}
{50\keV}\right)^{-3/4}n_{l,18}^{1/2}\eV,\cr
\end{align}
where $n_l=10^{18}n_{l,18}$. The lepton number density $n_l$ is roughly given by
\begin{align}\label{eq:MRelLeptons}
n_l&=\left(\frac{n_l}{n_\gamma}\right)n_\gamma \sim \left(\frac{n_l}{n_\gamma}\right)\frac{m_p\bt_s^2c^2}{3T}n_u\cr &\sim
0.6\times 10^{18}\left(\frac{10n_l}{n_\gamma}\right)\left(\frac{T}{50\keV}\right)^{-1}\bt_s^2 n_{15}\cm^{-3},\cr
\end{align}
yielding
\begin{equation}\label{eq:AbsorptionLMRel}
\Lm_a=\ln(\frac{T}{h\nu_a})\approx 11-\frac12 \ln(n_{15}),
%+\ln\left[\left(\frac{T}{50\keV}\right)^{9/4}\left(\frac{30n_l}{n_\gamma}\right)^{-1/2}\right],
\end{equation}
where we used $g_{ff,ll}\sim 10,n_l/n_\gamma\sim10$ and $T\sim 50\keV$ (the result is not sensitive to the specific
values adopted for these parameters).

The Compton $y$ parameter is
\begin{equation}\label{eq:ComptonyRel}
y=(4\hat T+16\hat T^2)\bt_d^{-2}> 40 (T/50\keV) \bt_{d,-1}^{-2}.
\end{equation}
Equations~\eqref{eq:AbsorptionLMRel} and \eqref{eq:ComptonyRel} imply that for $T\sim0.1m_ec^2$, the Compton $y$ parameters is large enough, $y>\Lm_a$, to allow all the photons above the absorption frequency, $h\nu_a$, to be comptonized (photons are upscattered by a factor of $e^y$ in energy moving all photons above $h\nu_a$ to the Comptonized energies $3T=3e^{\Lm_a}h\nu_a$) .
This implies, in turn, that photon production at all frequencies above $h\nu_a$ contribute to the effective generation rate and $\Lm_{\text{eff}}=\Lm_a$.

Let us compare now the photon-lepton ratio derived for the balance of photon production and diffusion, to the ratio derived from pair production equilibrium. For large chemical potential, $\mu_{\text{ch}}\gg T$ pair production equilibrium gives
\begin{equation}
\frac{n_{l}}{n_\gamma}|_{\text{eq}}=\int_0^{\infty}dxx^2e^{-\sqrt{x^2+\hat T^{-2}}}=\frac{K_2(\hat T^{-1})}{\hat T^2}
\end{equation}
where $K_2$ is the modified bessel function of the second kind of order 2. At the temperature range $60\keV<T<90\keV$
this ratio changes by an order of magnitude: ${n_\gamma}/n_{l}|_{\text{eq}}(T=60\keV)\approx130$ while
${n_\gamma}/n_{l}|_{\text{eq}}(T=90\keV)\approx13$. Comparing these values to the result given by equation
$\eqref{eq:PhotonGenerationRel}$, we conclude that at $\bt_d=0.1$ the temperature must be above $60\keV$.

\subsection{The relativistic limit $\bt_d=1/3$}
\label{sec:bs1}

In the regime $200\keV<T<m_ec^2$, the pair production equilibrium is approximately given by
\begin{equation}\label{eq:EquilibriumHighT}
n_\gamma/n_l\approx 0.5 m_ec^2/T.
\end{equation}

The number of leptons is now given by (see Eq. \eqref{eq:MRelLeptons}),
\begin{align} n_l&\sim
\left(\frac{n_l}{n_\gamma}\right)\frac{\Gamma_s^2 m_p c^2}{3T}n_u\cr &\sim 0.6\times
10^{18}\left(\frac{n_l}{n_\gamma}\right)\left(\frac{T}{m_e c^2}\right)^{-1}\Gamma_s^2 n_{15}\cm^{-3},\cr
\end{align}
and the absorption photon energy by
\begin{align}\label{eq:AbsorptionCutoffRel2}
h\nu_{a}\sim 0.1 \left(\frac{g_{\text{ll,ff}}}{10}\right)^{1/2}\left(\frac{T} {m_ec^2}\right)^{-3/4}n_{l,18}^{1/2}\eV,\cr
\end{align}
yielding
\begin{equation}\label{eq:AbsorptionLRel}
\Lm_a=\ln(\frac{T}{h\nu_a})\approx 15-\frac12 \ln(n_{15})-\ln(\Gamma_s),
%+\ln\left[\left(\frac{T}{50\keV}\right)^{9/4}\left(\frac{30n_l}{n_\gamma}\right)^{-1/2}\right],
\end{equation}
where we used $g_{ff,ll}\sim 10,n_l/n_\gamma\sim1$ and $T\sim m_ec^2$ (the result is not sensitive to the specific
values adopted for these parameters).

We can rewrite equation \eqref{eq:PhotonGenerationRel} for $\bt_d\approx 1/3$ and ${60\keV<T<m_ec^2}$ as
\begin{equation}\label{eq:PhotonGenerationRel2}
\frac{n_\gamma}{n_l}\approx 2.5 \left(\frac{\Lm_{\text{eff}}}{15}\right)^2(3\bt_{d})^{-2}.
\end{equation}

Comparing equations \eqref{eq:PhotonGenerationRel2} and \eqref{eq:EquilibriumHighT}, we see that if ${T\gtrsim 200\keV}$
there would be too many photons generated per lepton. At these high temperatures, the Compton $y$ parameter is large and
radiative compton emission is negligible. At temperatures $T\gtrsim 100\keV$ KN corrections to the Compton cross-section
of photons with energy $h\nu\sim 3T$ become significant. The possible effect that KN corrections have is to lower the temperature due to the larger photon generation depth.
This in turn would suppress the KN corrections. Hence, KN corrections cannot change the conclusion that $T\lesssim 200\keV$.
We conclude that for $\bt_d>0.1$, $60\keV<T\lesssim 200\keV$.

\section{Supernova Shock Breakouts}\label{sec:Breakouts}
Once an expanding radiation mediated shock reaches a point where the residual optical depth is of order $\tau\sim
\bt^{-1}$, whether at the outer shell of a star's envelope or in the surrounding wind (in case of an optically thick
wind), the radiation escapes, and the shock no longer sustains itself. In this section we summarize the relations between
the stellar/wind parameters and the velocity, energy and duration of the resulting shock-breakout radiation outburst. We
first derive general relations between the energy, velocity, radius and duration of the outburst in
\sref{sec:BreakoutGeneral}. Next, we discuss in \sref{sec:BreakoutDynamics} the dynamics of shock acceleration near the
stellar surface, and its evolution within the wind, and derive characteristic breakout velocities. The X-ray
characteristics of the breakout are discussed in \S~\ref{sec:BreakoutEmission}, based on the results of
\S~\ref{sec:NRRMS}, \S~\ref{sec:RRMS}, and \S~\ref{sec:BreakoutDynamics}. Finally, the implications of our results to the
interpretation of recent X-ray outbursts associated with SNe (XRO080109/SN2008D, XRF060218/SN2006aj), and possibly to
other SN-associated outbursts (XRFs/GRBs), are discussed in \S~\ref{sec:observations}. In this section, we restrict our discussion to non-relativistic shocks $\bt_s\lesssim0.6$ and non relativistic temperatures $T_s\lesssim 50 \keV$, corresponding to photon energies $h\nu\lesssim 150 \keV$, where pair creation can be neglected.

\subsection{A general relation between the breakout energy, velocity, and radius}\label{sec:BreakoutGeneral}
Consider a shock with velocity $\bt_s\gtrsim 0.1$ approaching the photosphere of a spherical mass distribution. The break-out will occur once the optical depth down to the photosphere is $\tau\sim \bt^{-1} \lesssim 10$.

The energy carried by the photons in a shell of shocked material with a width of the order of the shock width is roughly
given by \citep[e.g.][]{Waxman07}
\begin{equation}\label{eq:Edembt}
\de E_{\gamma}= f_E\de m \bt_s^2c^2/2,
\end{equation}
where $\de m$ is the mass of the shell and $f_E$ is a coefficient that depends on the velocity and on the geometry. For a shocked shell in the downstream of a non-relativistic planar shock, $f_E=36/49$. $f_E$ is of order unity for non relativistic and mildly relativistic shocks in spherical geometry, as long as the distance to the photosphere is not much larger than the radius.
The mass of a shell that has an optical depth $\tau$ is roughly given by
\begin{equation}
\de m\sim 4\pi R^2\tau\kappa^{-1}\approx 10^{26} R_{12}^2 (\kappa/\kappa_T)^{-1}\tau_{0.5}\gr,
\end{equation}
where $\kappa_T=\sig_T/m_p\sim 0.4\cm^2\gr^{-1}$, $\tau=10^{0.5}\tau_{0.5}$ and ${R=10^{12}R_{12}\cm}$. Assuming that the
energy of shocked plasma with optical depth $\tau\sim 3\tau_{0.5}$ is emitted, we find that there is a simple relation
between the released energy, breakout radius and shock velocity at breakout,
\begin{equation}\label{eq:EVR}
\de E \sim 0.5\de m (\bt_s c)^2\sim 4\times 10^{46} \bt_s^2R_{12}^2 (\kappa/\kappa_T)^{-1}\tau_{0.5}\erg.
\end{equation}

Once a significant fraction of the energy in the shock region is emitted, the steady state shock solution is no longer applicable.
Note that before the solution becomes invalid, a significant fraction of the energy in the shock region must be emitted.
We thus expect that photons with energies of the order of $3T_s$ carrying an energy similar to the energy estimated in
\eqref{eq:EVR} are inevitably emitted.

\subsection{Breakout Dynamics}\label{sec:BreakoutDynamics}
In this subsection we discuss the dynamics of shock acceleration near the stellar surface, and its evolution within the
wind, and derive characteristic breakout velocities. Many of the relations derived can be found in the literature
\citep[e.g.][]{Matzner99}.

At the outer layers of stars, where the enclosed mass, $M$, is approximately constant, $M=M_*=const$, and the composition
is approximately uniform, the density distribution is a power-law in the distance from the edge of the star,
\begin{equation}
\rho=\rho_1 \delta^{n},
\end{equation}
where $R_*$ is the stellar radius and
\begin{equation}
\delta=(R_*-r)/r.
\end{equation}
This relation is exact if we assume $M=M_*$ and a polytropic equation of state $p=k\rho^{\gamma_n}$, in which case $n=(\gamma_n-1)^{-1}$ and
\begin{equation}
\rho_1=\left(\frac{\gamma_n-1}{k\gamma}\frac{GM}{R_*}\right)^{n}.
\end{equation}
For an efficiently convective envelope, the entropy is constant. Assuming an adiabatic index $\gamma_n=5/3$, we have $n=3/2$.
For a radiative envelope, with radius independent luminosity ${L=L_{*}\ll L_{\text{Edd}}}$ and mass $M=M_*$, we have $p_{\gamma}=(L_{*}/L_{Edd})p$, where $L_{\text{Edd}}=4\pi GMc/\kappa$ is the Eddington luminosity limit, resulting in a polytropic equation of state $p\propto \rho^{4/3}$, or $n=3$. The optical depth from the edge of the star is given by
\begin{equation}\label{eq:OpticalDepth}
\tau=\int_r^{R_*}\rho\kappa dr\propto \delta^{n+1}.
\end{equation}
The density at the outer edge is given by
\begin{equation}
\rho=(\rho_1)^{\inv{n+1}}\left(\frac{(n+1)\tau}{\kappa R_*}\right)^{n/(n+1)}.
\end{equation}
Denoting
\begin{equation}
\rho_1=f_\rho \frac{M_{\text{ej}}}{\frac{4\pi}{3}R_*^3},
\end{equation}
where $M_{\text{ej}}$ is the mass of the ejected envelope, we find
\begin{equation}\label{eq:rhot}
\rho\approx 2.0\times 10^{-9}f_{\rho}^{1/4}(M_{\text{ej}}/M_\odot)^{1/4}(\kappa/\kappa_T)^{-3/4}R_{12}^{-3/2}\tau_{0.5}^{3/4}\gr\cm^{-3}
\end{equation}
for $n=3$, and
\begin{equation}
\rho\approx 2.8\times 10^{-10}f_{\rho}^{2/5}(M_{\text{ej}}/M_\odot)^{2/5}(\kappa/\kappa_T)^{-3/5}R_{13}^{-9/5}\tau_{0.5}^{3/5}\gr\cm^{-3}
\end{equation}
for $n=3/2$.

As the shock wave approaches the stellar surface, it accelerates down the density gradient. In planar geometry, with density vanishing at $x=x_0$ following $\rho\propto[(x_0-x)/x_0]^n=\delta^n$, the flow approaches a self-similar behavior as the shock velocity diverges \citep{Gandel'Man56,Sakurai60}. In this limit
\begin{equation}\label{eq:vsproptau}
\vt_s\propto\delta^{-\bt_1n}\propto \tau^{-\lambda_{\vt}},
\end{equation}
where $\lambda_{\vt}=\bt_1n/(n+1)$, and $\bt_1\simeq0.19$ for $\gamma=4/3$ and $n=3,3/2$. The planar self-similar
solution provides a good approximation for the shock behavior near the surface of the star, where the thickness of the
layer lying ahead of the shock is small compared to the radius, i.e. for $\delta\ll1$ \citep[e.g.][]{Matzner99}. Denoting
the mass $M_*-M(r)$ of the shell lying ahead of radius $r$, at $\delta<(R_*-r)/R_*$, by $m=M_*-M(r)$, and noting that
near the surface the optical depth $\tau$ is $\tau\propto m$, equation~\eqref{eq:vsproptau} may be written as
\begin{equation}\label{eq:vm}
\vt_s\propto m^{-\lm_v}
\end{equation}
with $\lm_v=\bt_1n/(n+1)$.
The kinetic energy carried by the material having velocity greater than some $\vt_s$ scales with the velocity as
\begin{equation}\label{eq:EVstar}
E(>\vt_s)\propto \vt_s^{-\de_{\vt}},
\end{equation}
where
\begin{equation}
\de_{\vt}=\frac{1+\inv{n}-2\bt_1}{\bt_1}\approx 5, 7
\end{equation}
for $n=3$ and $n=3/2$ respectively.

\begin{figure}[h]
\epsscale{1} \plotone{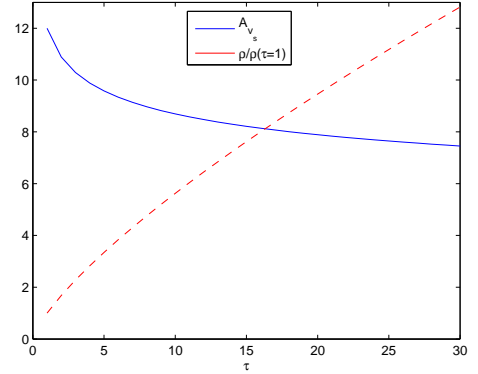}
\caption{\label{fig:RhoVBSG} Shock velocity amplification, and density typical of the edge of a Blue Super-Giant (n=3,$M_{SeS}=M_{\odot}$, $R=10^{12}\cm$), as a function of the optical depth from the edge of the star as given by Eqs. \eqref{eq:vst} and \eqref{eq:rhot}. The density is normalized to the value at $\tau=1$.}
\end{figure}

The velocity of the shock at an optical depth $\tau$ is larger than the typical bulk envelope velocity, $\vt_{\rm ej,b.}\approx\sqrt{E_{\rm ej}/M_{\rm ej.}}$ where $E_{\rm ej}$ is the energy deposited in the envelope, by a factor
\begin{equation}
A_{\vt_s}\approx \left(\frac{M_{\text{Ses}}\kappa}{4\pi R_*^2}\tau\right)^{\lambda_{\vt}},
\end{equation}
where $M_{\text{Ses}}$ is the mass of the region in which the planar solution is applicable.
For $n=3$ we have $\lambda_{\vt}\approx 0.14$, while for $n=3/2$ we have $\lambda_{\vt}\approx 0.11$.
The velocity amplification factor for $n=3$ is
\begin{equation}\label{eq:vst}
A_{\vt_s}\approx 11 (M_{\text{Ses}}/M_\odot)^{0.14}(\kappa/\kappa_T)^{0.14}R_{12}^{-0.28}\tau_{0.5}^{-0.14},
\end{equation}
and for $n=3/2$ it is
\begin{equation}
A_{\vt_s}\approx 4 (M_{\text{Ses}}/M_\odot)^{0.11}(\kappa/\kappa_T)^{0.11}R_{13}^{-0.23}\tau_{0.5}^{-0.11}.
\end{equation}
The density and velocity profiles of the outer envelope are illustrated in figure \ref{fig:RhoVBSG}.

If there is an optically thick wind surrounding the star, breakout occurs within the wind.
Assuming a wind with a constant mass loss rate $\dot M$ and velocity $\vt_w$, $\rho_w=\dot M/4\pi \vt_{w}R^{2}$,
the optical depth between $R$ and infinity is related to the mass $m_w(<R)$ by $\tau(>R)=m_w(<R)\kappa/4\pi R^2$ (for $R\gg R_*$), and breakout is expected to occur at a radius
\begin{align}
R_{br}=\frac{\dot M}{\vt_w}\frac{\kappa}{4\pi\tau}
\sim 0.6\times 10^{11}\left(\frac{\dot M_{-5}}{\vt_{w,8}}\right)(\kappa/\kappa_T)\tau_{0.5}^{-1}\cm.
\end{align}
At early time, the velocity $\vt_{s,w}$ of the shock driven into the wind at radius $R$ is approximately given by the
velocity of the fastest shell of the ejecta. Since the final velocity reached by a fluid element shocked at $\vt_s$,
following its acceleration due to the post-shock adiabatic expansion, is $\vt_f\simeq2\vt_s$ \citep{Matzner99}, the
velocity of the shock driven into the wind is initially $\approx 2 A_{\vt_s}\vt_{\text{ej}}$. At later time, when the
accumulated wind mass begins to decelerate the ejecta, the shock velocity is approximately obtained by equating the mass
of the shocked wind, $m_w(<R)$, with the mass of the part of the ejecta that was accelerated to velocity $\vt>\vt_{s,w}$.
Using eq.~(\ref{eq:vm}) we have $m_w(<R)\approx M_{\rm ej}(\vt_{s,w}/2\vt_{\rm ej})^{-\lambda_\vt}$ or
$\vt_{s,w}/\vt_{\rm ej}\approx 2[M_{\rm ej}/m_w(<R)]^{\lambda_\vt}$. This gives:
\begin{equation}\label{eq_AvwB}
  \frac{\vt_{s,w}}{\vt_{\rm ej}}=\min\left[2A_{\vt_s}, 30\left(\frac{M_{\rm ej}}{M_\odot}\right)^{1/7}\left(\frac{\dot M_{-5}}{\vt_{w,8}}\right)^{-1/7}R_{12}^{-1/7} \right]
\end{equation}
for BSG ($n=3$) and
\begin{equation}\label{eq_AvwR}
  \frac{\vt_{s,w}}{\vt_{\rm ej}}=\min\left[2A_{\vt_s}, 13\left(\frac{M_{\rm ej}}{M_\odot}\right)^{1/9}\left(\frac{\dot M_{-5}}{\vt_{w,8}}\right)^{-1/9}R_{13}^{-1/7} \right]
\end{equation}
for RSG ($n=3/2$).

The assumption that the shock velocity in the wind is equal to the velocity of the shocked ejecta is correct as long as
the density encountered by the reverse shock in the ejecta is much larger than the density encountered by the forward
shock in the wind.

 Assuming typical bulk envelope velocities of $\vt_{\text{ej,b}}\approx\sqrt{E_{\rm ej}/M_{\rm ej.}}\sim
3-10\times 10^8\cm/{\rm s}$, or $\bt_{\text{ej, b}}\sim 0.01-0.03$, the typical shock velocities at the last few optical
depths of BSGs and WR stars may reach values of $\bt_s\gtrsim 0.2$, for which temperatures exceeding $T\sim 10\keV$ may
be reached in the shock transition region. The shock velocities in the wind may be up to twice higher than the maximal
shock velocity achieved within the star.

\subsection{Breakout X-ray characteristics}\label{sec:BreakoutEmission}

Let us consider the radiation that may be emitted during the break-out of a shock from the stellar surface. We consider the emission during a time interval $R_*/c$ following the time at which the shock reached the surface (as long as the star is not resolved, any emission will be spread in time over this time scale). On similar time scales, the outer part of the star expands significantly, over a distance $\bt_f R_*\sim R_*$, where $\beta_f\approx2\beta_s$ is the final velocity reached by a fluid element shocked at $\beta_s$ \citep[following its acceleration due to the post-shock adiabatic expansion, e.g.][]{Matzner99}.
On the same time scale, $R_*/c$, photons that originated at an optical depth $\tau_{\text{esc}}\sim\delta^{-1/2}(\tau=1)$ are capable of escaping. Using equation \eqref{eq:OpticalDepth} we have
\begin{equation}
\delta(\tau=1)\sim\left(\frac{M_{\text{ej}}\kappa}{4\pi R_*^2}\right)^{-\inv{n+1}}.
\end{equation}
This implies
\begin{equation}
\tau_{\text{esc}}\sim 10(M_{\text{ej}}/M_\odot)^{1/8}(\kappa/\kappa_T)^{1/8}R_{12}^{-1/4}
\end{equation}
for BSG typical parameters, and
\begin{equation}
\tau_{\text{esc}}\sim 15(M_{\text{ej}}/M_\odot)^{1/5}(\kappa/\kappa_T)^{1/5}R_{13}^{-2/5}
\end{equation}
for RSG parameters. For $1\gtrsim\bt_s\gtrsim0.1$, these values are not much larger than the shock widths, $\bt_s^{-1}$,
and a non negligible fraction, $\bt_s^{-1}/\tau_{esc}$, of the photons in these regions are likely to escape over $\sim
R_*/c$. The amount of energy emitted during $R_*/c$ is thus not much larger than the estimate given in
eq. \eqref{eq:EVR}.

Note that the radiation observed at a given time is the sum of radiation emitted from different positions of the star at
different stages of the breakout (up to differences of $R_*/c$). Photons that originated from shocked material in deeper
layers, $\tau\gg \bt_s^{-1}$, will have characteristic energies which are lower than the photon energies at the
transition region $h\nu\sim 3 T_s$ due to photon production and possible adiabatic cooling. Therefore we expect to see a
spectrum which is a sum of Wein-like spectra with temperatures reaching the tens of keV shock temperatures $T_s$ and
starting from lower energies.

Let us consider next breakouts that take place within an optically thick wind. In this case, the optical depth of the
shocked wind material, which is compressed to a thin shell, is comparable to the optical depth of the wind ahead of it.
Thus, we expect all photons contained within this region to escape on a time scale similar to $R/\vs$, the time scale on
which the optical depth changes significantly. Again, the spectrum would be composed of Wein-like spectra with
temperatures reaching $T_s$.

Finally, the following point should be stressed. As the shock approaches an optical depth of $1/\bt$ from the surface, the shock structure deviates from the steady state solution due to the escape of energy from the system and due to the steep density gradient. Note, that at this stage the shock propagates within a density profile which varies on a length scale comparable to the shock thickness. As for the effect of energy escape, a significant fraction of the energy in the shock transition region must be emitted before the structure is changed and thus the steady state estimate for the photon energies will correctly describe the photons carrying most of the escaping energy. It is reasonable to expect that the effect of the density gradient on the temperature will be of order unity for fast shocks, as the pre-shock density across a shock width changes only by a factor of few- by $\sim 1/\bt$ in the case of stellar surface [see Eqs. \eqref{eq:rhot}] and by $\sim 1$ in the case of a wind. An exact calculation of the emerging spectrum requires a time dependent calculation and is beyond the scope of this paper.

\subsection{Implications to recently observed SN X-ray outbursts}
\label{sec:observations}

It was long been suggested that a strong outburst of Ultraviolet or X-ray radiation may be observable when the RMS reaches the edge of the star \citep[e.g.][]{Colgate74b,Falk78,Klein78,Ensman92,Matzner99,Blinnikov00}. These outbursts carry important information about the progenitor star \citep{Arnett77} including a direct measure of the star's radius. During the past two years, the wide field X-ray detectors on board the Swift satellite detected luminous X-ray outbursts preceding a SN explosion in two cases, SN2006aj and SN2008D \citep{Campana06,Soderberg08}. Analysis of the later optical SN emission revealed that both were of type Ib/c, probably produced by compact (WR) progenitor stars \citep{Pian06_SN_aj,Mazzali06_prog_aj,Modjaz06_aj_SN_analysis,Maeda07_nucleo_aj_nomoto,Mazzali08_08D_SN,
Malesani09_08d_spectr,Tanaka08_08D_prog_nomoto}. While some authors argue that the X-ray outbursts are produced by shock breakouts \citep{Campana06,Waxman07,Soderberg08}, others argue that the spectral properties of the X-ray bursts rule out a breakout interpretation, and imply the existence of relativistic energetic jets penetrating through the stellar mantle/envelope \citep{Soderberg06_rel_jet_aj,Fan06_aj,Ghisellini07_aj,Li07_aj,Mazzali08_08D_SN,Li08_08D_XRF}
 \citep[some argue that the breakout interpretation holds for SN2008D, but not for SN2006aj][]{Chevalier08}.

The main challenge raised for the breakout interpretation \citep[e.g.][]{Mazzali08_08D_SN} was that in both cases,
SN2006aj and SN2008D, the observed X-ray flash had a non-thermal spectrum extending to $\gtrsim10\keV$, in contrast with the sub-$\keV$ thermal temperatures expected \citep{Matzner99} \citep[Note, that while thermal components with low temperatures or fluxes cannot be ruled out, e.g. ][ such thermal components carry only a small fraction of the observed flux]{Mazzali08_08D_SN, Modjaz08_08D}. However, the analysis presented here implies that fast, $\bt_s>0.2$, breakouts may be expected for compact, BSG or WR progenitors, and that for such fast breakouts the X-ray outburst spectrum may naturally extend to 10's or 100's of keV.

Let us discuss in some more detail the possible interpretation of the X-ray outbursts associated with  SN2006aj and SN2008D as breakouts. In the case of SN2008D, both the X-ray outburst and early UV/O emission that followed it are consistent with a breakout interpretation \citep[see][for detailed discussion]{Soderberg08}. The X-ray burst energy and duration are consistent with the relation of eq.~\ref{eq:EVR}, and the UV/O emission is consistent with that expected from the post-shock expansion of the envelope. Moreover, the non-thermal X-ray and radio emission that follow the X-ray outburst are consistent with those expected from a shock driven into a wind by the fast shell that was accelerated by the radiation mediated shock, for wind density and shell velocity, $\vt/c\sim1/4$, and energy, $E\sim10^{47}$~erg, which are inferred from the X-ray outburst. The fast velocity inferred for the late   radio and X-ray emission is consistent with that required to account for the non-thermal spectrum.

The case of SN2006aj is more complicated. The X-ray burst energy and temperature are consistent with a mildly
relativistic, $\vt/c\sim0.8$, shock breakout from a wind surrounding the star, the UV/O emission is broadly consistent
with that expected from the expanding envelope, and the non-thermal X-ray emission that follows the X-ray outburst is
consistent with that expected from the shock driven into the wind by the fast shell that was accelerated by the radiation
mediated shock \citep{Campana06,Waxman07}. However, the duration of the X-ray outburst is larger than expected, the UV/O
emission is not as well fit by the model as in the case of 2008D, and the non-thermal radio emission is higher than
expected from the wind-shell interaction. Several authors \citep{Campana06,Waxman07} have suggested that the deviations
from the simple breakout model are due to a highly non-spherical explosion. Some support for the non spherical nature of
the explosion was later obtained the polarization measurements of \citet{Gorosabel06_pol_aj} \citep[see,
however,][]{Mazzali07_no_pol_aj}. Other authors \citep{Soderberg06_rel_jet_aj,Ghisellini07_aj,Fan06_aj,Li07} have argued
that a relativistic jet is required to account for the observations. We believe that an analysis of the modifications to
the simple spherical model, introduced by a highly non-spherical breakout, is required to make progress towards
discriminating between the two scenarios.

There is an additional important point that should be clarified in the context of SN2006aj. Under the shock breakout hypothesis, the energy and velocity of the accelerated shell that is responsible for the X-ray emission are $\vt/c\sim0.8$ and $E\sim10^{49.5}$~erg. This is similar to the parameters of the fast expanding shell inferred to be ejected by SN1998bw (associated with GRB080425), based on long term radio and X-ray emission, which are interpreted as due to interaction with a low density wind \citep{Kulkarni98,WnL99,LinC99,Waxman_bw1,Waxman_bw2}. Long term radio observations strongly disfavor the existence of an energetic, $10^{51}$~erg, relativistic jet associated with SN1998bw \citep{Soderberg04_bw_nojet}. The similarity of SN1998bw \& SN2006aj may   therefore suggest that such a jet is not present also in the case of SN2006aj. However, the large energy deposited by the explosion in a mildly relativistic shell, $\vt/c\sim0.8$, is a challenge in itself: The acceleration of the supernova shock near the edge of the star is typically expected to deposit only $\sim10^{46}$~ergs in such fast shells \citep{Tan01}, as can be inferred, e.g, from eq.~(\ref{eq:EVstar}), which implies $E(>\vt_s)/E_{\rm ej.}\sim (\vt_s/\vt_{\rm ej.,b})^{-\de_{\vt}}$.

\section{Discussion}\label{sec:Discussion}

We presented a simple analytic model for non-relativistic (\sref{sec:NRRMS}) and relativistic (\sref{sec:RRMS}) radiation mediated shocks. At shock velocities $\bt_s=\vt_s/c\gtrsim 0.1 (n/10^{15}\cm^{-3})^{1/30}$ [Eq. \eqref{eq:btNeq}], where $n$ is the upstream density, the plasma is far from thermal equilibrium within the transition region, where most of the deceleration takes place, since the plasma does not have enough time to generate the downstream black-body photon density $n_{d,\gamma}\sim a_{BB}T_d^3/3$. The electrons (and positrons) are heated in this region to temperatures $T_s$ significantly exceeding the far downstream temperature, $T_s\gg T_d$, and the photons are in Compton equilibrium with the electrons (and positrons). The transition temperature, $T_s$, is independent of the upstream density: It is given by $\beta_s\approx0.2(T_s/10\keV)^{1/8}$ for $\bt_s\lesssim 0.2$ [eq.~\eqref{eq:btOfT}]; At velocities $\bt_s\gtrsim0.6$ ($\bt_d \gtrsim 0.1$), where the plasma is dominated by electron-positron pairs in pair production equilibrium, the temperature is constrained to the range $60\keV\lesssim T_s \lesssim 200\keV$. The thermalization length, the distance behind the transition region over which the plasma thermalizes and reaches the downstream temperature $T_d$, is $\sim(T_s/T_d)^{1/2}$ times larger than the transition (deceleration) width, $\sim1/\bt_s \sigma_T n$. Our simple model estimates are in agreement with the results of exact numerical relativistic calculations, which will be presented in a follow-up paper \citep{Budnik10}.

We showed in section \sref{sec:Breakouts} that radiation mediated shocks breaking out of the stellar envelopes of BSGs
and Wolf-rayet stars are likely to reach velocities $\bt_s>0.2$. We thus expect that for reasonable stellar parameters,
the spectrum emitted during supernovae shock breakouts from BSGs and Wolf-Rayet stars may include a hard component with
photon energies reaching tens or even hundreds of $\keV$. This implies that core-collapse SNe produced by BSGs/WR stars
may be searched for by using hard X-ray (wide field) detectors. The detection rate of such events is significantly
different than is inferred assuming a thermal emission spectrum \citep[e.g.][]{Calzavara04}, due to both the modification
of the intrinsic spectrum and to the reduced absorption of high energy photons. A quantitative estimate of the increased
detection rate is beyond the scope of this paper. As the escaping photon energies are highly sensitive to the shock
velocity, the rate is sensitive to the unknown high-end part of the distribution of the shock velocities during
breakouts.

We have argued in \S~\ref{sec:observations} that the X-ray outburst XRO080109 associated with SN2008D is most likely due
to a fast breakout: The energy, $\sim10^{47}$~erg, duration, $\sim 30$~s, and fast ejecta velocity $\beta_s\simeq1/4$,
are all consistent with expected breakout parameters, and the hard X-ray spectrum is a natural consequence of the high
velocity, which is independently inferred also from later X-ray and radio observations. Our analysis shows that the
spectrum of the X-ray flash XRF060218, associated with SN2006aj, might also be explained as a fast breakout. However, the
breakout interpretation of this event is challenged by the long duration of the X-ray flash, and by the high energy,
$E\sim10^{49.5}$~erg, deposited in this explosion in mildly relativistic, $\vt/c\sim0.8$, ejecta (see detailed discussion
in \S~\ref{sec:observations}).

\citet{Wang07} have suggested, based on the breakout interpretation of the X-ray outbursts of SN2006aj and SN2008D, that
all the low-luminosity Gamma-ray bursts/X-ray flushes associated with SNe, which have smooth light curves and spectra not
extending beyond few 100~keV (like those associated with SN1998bw, SN2003lw, SN2006aj), are due to shock breakouts, and
do not require the existence of energetic highly relativistic jets. The present analysis, which demonstrates that fast
breakouts may indeed produce non-thermal spectra extending to 100's of keV, supports the viability of the breakout
interpretation of low-luminosity Gamma-ray bursts/X-ray flushes associated with SNe. A major challenge for such a
scenario is constituted by the requirement of large energy deposition, $E\sim10^{49.5}$~erg, in the fastest, mildly
relativistic ($\vt/c\sim0.8$), part of the expanding ejecta.

\acknowledgements This research was partially supported by Minerva, ISF and AEC grants.

\appendix
\section{Solving the Non-Relativistic RMS equations}\label{sec:DetailedSol}
We describe below our numerical solution of the NR RMS equations. The derivation given below offers several advantages compared to the derivation given by \citet{Weaver76}: The equations are written in a dimensionless form, making the dependence of shock structure on parameters (in particular, on plasma density) explicit, and the diffusion equation is solved analytically, making the numerical integration procedure much simpler.

We use the following approximations:
\begin{itemize}
\item We neglect the pressure of the electrons and protons;
\item We assume that the radiation and the electrons are in Compton- equilibrium with equal temperatures $T_\gamma=T_e$;
\item We neglect the corrections to the Compton scattering cross section and assume it to be equal to the Thompson cross section;
\item We assume that the process generating the photons is Bremsstrahlung emission, as given by Eq. \eqref{eq:QgammaBr};
\item We assume that the photon distribution is described by the diffusion equation.
\end{itemize}

Under these assumptions, the equations determining the steady state shock profile are
conservation of proton flux,
\begin{equation}\label{eq:AConsPar}
n_p\bt=n_u\bt_s,
\end{equation}
conservation of momentum flux,
\begin{equation}\label{eq:AConsMom}
p_\gamma=n_u\bt_s(\bt_s-\bt)m_pc^2,
\end{equation}
and conservation of energy flux,
\begin{equation}\label{eq:AConsEng}
(e_\gamma+p_\gamma)\bt c-\frac{c}{3n_p\sig_T}\frac{de_\gamma}{dx}=n_u\bt_s m_p\frac{\bt_s^2-\bt^2}2c^3.
\end{equation}
These equations, together with the relation
\begin{equation}\label{eq:AEqState}
e_\gamma=3p_\gamma,
\end{equation}
form a closed set of equations for the velocity $\bt$, particle number density, $n_p=n_e$, and photon pressure, $p_\gamma$, and energy, $e_\gamma$.

$e_\gamma$, $p_\gamma$ and $n$ may be eliminated from eqs.~\eqref{eq:AConsPar}-\eqref{eq:AEqState} to obtain an equation for the velocity,
\begin{equation}\label{eq:Abt}
\frac{d\tilde\bt}{d\tilde x}=-\frac{(7\tilde\bt-1)(1-\tilde\bt)}{6\tilde\bt},
\end{equation}
where we introduced the dimensionless variables $\tilde x=3\sig_Tn_u\bt_s x$ and $\tilde\bt=\bt/\bt_s$.
Equation \eqref{eq:Abt} can be solved analytically \citep[e.g.][]{Weaver76},
\begin{equation}\label{eq:AanalyticSolution}
\tilde x=\frac{1}{7}\ln\left[\frac{(1-\tilde\bt)^7}{(7\tilde\bt-1)}\right].
\end{equation}

The photon pressure is related to the temperature through
\begin{equation}\label{eq:ApnT}
p_\gamma=n_\gamma T,
\end{equation}
where $n_\gamma$ is the photon number density. $T$, or $n_\gamma$, are determined
by the photon diffusion eq.,
\begin{equation}\label{eq:ADiff}
\frac{dj_\gamma}{dx}=Q_{\gamma,\text{eff}},
\end{equation}
where
\begin{equation}\label{eq:photondiffusion}
j_\gamma=c\bt n_\gamma -\frac{c}{3n_p\sig_T}\frac{dn_\gamma}{dx}=c\bt\left(n_\gamma -\frac{1}{3n_u\bt_s\sig_T}\frac{dn_\gamma}{dx}\right)
\end{equation}
is the effective current of photons and the generation of photons is given by Eq. \eqref{eq:QgammaBr},
\begin{equation}\label{eq:AQgammaBr}
Q_{\gamma,\text{eff}}=\al_e n_pn_e\sig_T c\sqrt{\frac{m_ec^2}{T}}\Lm_{\text{eff}}g_{\text{eff}}f_{\text{abs}}.
\end{equation}
Here, $f_{\text{abs}}=1-n_\gamma/(a_{BB}T^3/3)$ is approximately the suppression factor due to absorption of the photons \citep{Weaver76}.
Equations \eqref{eq:ADiff}-\eqref{eq:ApnT} determine $T$ (and $n_\gamma$) for given $\bt$ and $p_\gamma$.

Defining dimensionless variables $\tilde T$ and $\tilde n_\gamma$ that satisfy
\begin{equation}\label{eq:ATnorm}
\frac{T}{m_ec^2}=\left(\frac{\vep}{m_ec^2}\right)^4 \left(\frac{m_e}{m_p}\right)^2 \al_e^{-2}\tilde T,
\end{equation}
and
\begin{align}\label{eq:Annorm}
n_\gamma=\left(\frac{\vep}{m_ec^2}\right)^{-3} \left(\frac{m_e}{m_p}\right)^{-2} \al_e^2n_u\tilde n_\gamma,
\end{align}
equation~\eqref{eq:ADiff} may be written as
\begin{equation}\label{eq:ADimlessDiff}
\frac{d\tilde j_{\gamma}}{d\tilde x}=\tilde Q_{\gamma},
\end{equation}
where
\begin{equation}\label{eq:ADimlessj}
\tilde j_{\gamma}=\tilde\bt(\tilde n_{\gamma}-\frac{d}{d\tilde x}\tilde n_{\gamma}),
\end{equation}
and
\begin{equation}\label{eq:AtildeQ}
\tilde Q=\Lm_{\text{eff}}g_{\text{eff}}f_{\text{abs}}\inv{6}\tilde\bt^{-2}\tilde T^{-1/2}.
\end{equation}
Using Eq. \eqref{eq:AConsMom} we have in addition
\begin{equation}\label{eq:AtildeTn}
\tilde T=\frac{2(1-\tilde\bt)}{\tilde n_{\gamma}}.
\end{equation}

The solution for $\tilde T$ and $\tilde n_\gamma$ is nearly independent of $\bt,\ n_p,\ T$, since these parameters appear in the equations only through the product $\Lm_{\text{eff}}g_{\text{eff}}f_{\text{abs}}$, which is weakly dependent on parameters. In particular, equation \eqref{eq:btOfT} satisfies the scaling given by Eqs.~\eqref{eq:ATnorm} and~\eqref{eq:Annorm}.

\begin{figure}[h]
\epsscale{0.7} \plotone{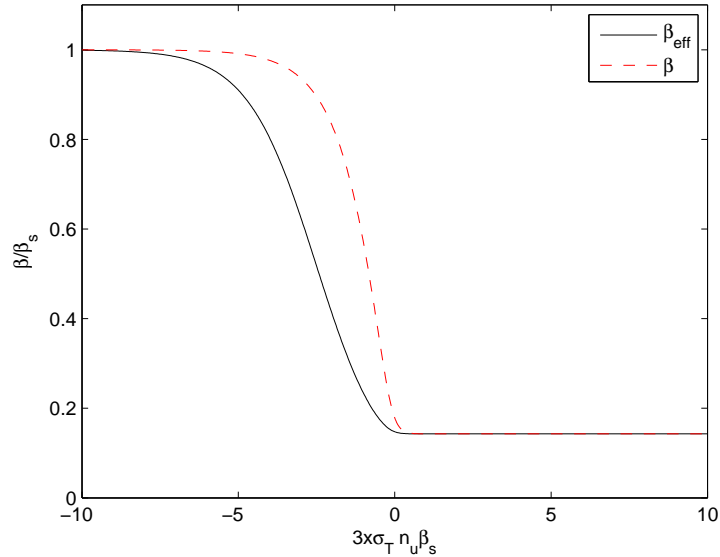} \caption{\label{fig:Abetaeff} Effective diffusion velocity defined by Eq.
\eqref{eq:Abteff} and calculated for the velocity profile given by Eq. \eqref{eq:AanalyticSolution}. The full black line
is the effective velocity and the red dotted line is the plasma velocity.}
\end{figure}

It is straight forward to show that the Green function, $G(\tilde x,\tilde x_0)$, for the diffusion equations \eqref{eq:ADimlessDiff} and \eqref{eq:ADimlessj},
defined by the relation
\begin{equation}\label{eq:AGreenDef}
\tilde n_\gamma(\tilde x)=\int_{-\infty}^{\infty}G(\tilde x,\tilde x_0)\tilde Q(\tilde x_0) d\tilde x_0,
\end{equation}
 is given by
\begin{equation}\label{eq:AGreenf}
G(\tilde x_0,\tilde x)=\left\{\begin{array}{cc}
\frac{e^{\tilde x-\tilde x_0}}{\tilde \bt_{\text{eff}}(\tilde x_0)}&~~~\tilde x<\tilde x_0\\
&\\
\frac{1}{\tilde \bt_{\text{eff}}(\tilde x)}&~~~\tilde x>\tilde x_0
\end{array}\right.,
\end{equation}
where,
\begin{equation}\label{eq:Abteff}
\inv{\tilde \bt_{\text{eff}}(\tilde x)}=\int_{\tilde x}^{\infty}e^{-(\tilde x'-\tilde x)}\inv{\tilde\bt(\tilde x')}d\tilde x'.
\end{equation}
The effective velocity $\tilde\beta_{\text{eff}}$, calculated for the velocity profile given by Eq. \eqref{eq:AanalyticSolution}, is shown in figure \ref{fig:Abetaeff}.

A numerical solution to equations \eqref{eq:ADimlessDiff}-\eqref{eq:AtildeTn} can be obtained by preforming iterations in which $\tilde Q_\gamma$ is calculated from $\tilde T$ using Eq. \eqref{eq:AtildeQ}, then $\tilde n_\gamma$ is calculated using Eq. \eqref{eq:AGreenDef}, and then a new value of $\tilde T$ is obtained using Eq. \eqref{eq:AtildeTn}. The solution presented in figure \ref{fig:Shock} was obtained for an interval $\tilde x_{\min}<\tilde x<\tilde x_{\max}$, ignoring photon contributions from outside of the interval, with $\tilde x_{\min}=\ln(m_e/m_p)-\ln(6)/7$ chosen as the point where $1-\bt=m_e/m_p$, and $x_{\max}=400$. The solution is largely unsensitive to the choice of the interval boundary.

\bibliographystyle{apj}

\begin{thebibliography}{21}
\expandafter\ifx\csname natexlab\endcsname\relax\def\natexlab#1{#1}\fi
%\bibitem[Blandford
%\& Payne(1981)]{Blandford81a} Blandford, R.~D., \& Payne, D.~G.\ 1981, \mnras, 194, 1033

\bibitem[Arnett(1977)]{Arnett77} Arnett, W.~D.\ 1977, Eighth
Texas Symposium on Relativistic Astrophysics, 302, 90

\bibitem[Becker(1988)]{Becker88} Becker, P.~A.\ 1988, \apj, 327,
772

\bibitem[Blandford
\& Payne(1981)]{Blandford81b} Blandford, R.~D., \& Payne, D.~G.\ 1981, \mnras, 194, 1041

\bibitem[Blinnikov et al.(2000)]{Blinnikov00} Blinnikov, S., Lundqvist, P., Bartunov, O., Nomoto, K., \& Iwamoto, K.\ 2000, \apj, 532, 1132

\bibitem[Bersier et al.(2006)]{Bersier06} Bersier, D., et al.\
2006, \apj, 643, 284

\bibitem[Budnik et al. (2010)]{Budnik10} Budnik, R., et al. \ 2010 in preperation.

\bibitem[Calzavara
\& Matzner(2004)]{Calzavara04} Calzavara, A.~J., \& Matzner, C.~D.\ 2004, \mnras, 351, 694


\bibitem[Campana et al.(2006)]{Campana06} Campana, S., et al.\
2006, \nat, 442, 1008

\bibitem[Cappa et al.(2004)]{Cappa04} Cappa, C., Goss, W.~M.,
\& van der Hucht, K.~A.\ 2004, \aj, 127, 2885

\bibitem[Chevalier
\& Klein(1979)]{Chevalier79} Chevalier, R.~A., \& Klein, R.~I.\ 1979, \apj, 234, 597

\bibitem[Chevalier
\& Fransson(2008)]{Chevalier08} Chevalier, R.~A., \& Fransson, C.\ 2008, \apjl, 683, L135


\bibitem[Colgate(1974)]{Colgate74} Colgate, S.~A.\ 1974, \apj,
187, 321

\bibitem[Colgate(1974)]{Colgate74b} Colgate, S.~A.\ 1974, \apj,
187, 333

\bibitem[Della Valle et
al.(2003)]{Della03} Della Valle, M., et al.\ 2003, \aap, 406, L33



\bibitem[Ensman
\& Burrows(1992)]{Ensman92} Ensman, L., \& Burrows, A.\ 1992, \apj, 393, 742


\bibitem[Falk(1978)]{Falk78} Falk, S.~W.\ 1978, \apjl, 225,
L133

\bibitem[Fan et al.(2006)]{Fan06_aj} Fan, Y.-Z., Piran, T., \& Xu, D.\ 2006, Journal of Cosmology and Astro-Particle Physics, 9, 13

\bibitem[Galama et al.(1998)]{Galama98} Galama, T.~J., et al.\
1998, \nat, 395, 670

\bibitem[Ghisellini et al.(2007)]{Ghisellini07_aj} Ghisellini, G., Ghirlanda, G., \& Tavecchio, F.\ 2007, \mnras, 375, L36

\bibitem[Gorosabel et al.(2006)]{Gorosabel06_pol_aj} Gorosabel, J., et al.\ 2006, \aap, 459, L33

\bibitem[Gandel'Man
\& Frank-Kamenetskii(1956)]{Gandel'Man56} Gandel'Man, G.~M., \& Frank-Kamenetskii, D.~A.\ 1956, Soviet Physics Doklady, 1, 223

\bibitem[Gorosabel et al.(2006)]{Gorosabel06_pol_aj} Gorosabel, J., et al.\ 2006, \aap, 459, L33

%\bibitem[Hutchinson(1987)]{Hutchinson88} Hutchinson, I. H., Principles of Plasma Diagnostics, Cambridge University Press

\bibitem[Klein
\& Chevalier(1978)]{Klein78} Klein, R.~I., \& Chevalier, R.~A.\ 1978, \apjl, 223, L109

\bibitem[Kulkarni et al.(1998)]{Kulkarni98} Kulkarni, S.~R., et al.\ 1998, \nat, 395, 663

\bibitem[Levinson
\& Bromberg(2008)]{Levinson08} Levinson, A., \& Bromberg, O.\ 2008, Physical Review Letters, 100, 131101

\bibitem[Li(2007)]{Li07_aj} Li, L.-X.\ 2007, \mnras, 375, 240

\bibitem[Li(2008)]{Li08_08D_XRF} Li, L.-X.\ 2008, \mnras, 388, 603

\bibitem[Li et al.(2007)]{Li07} Li, W., Wang, X., Van Dyk, S.~D., Cuillandre, J.-C., Foley, R.~J., \& Filippenko, A.~V.\ 2007, \apj, 661, 1013

\bibitem[Li \& Chevalier(1999)]{LinC99} Li, Z.-Y., \& Chevalier, R.~A.\ 1999, \apj, 526, 716

\bibitem[Lyubarskii
\& Sunyaev(1982)]{Lyubarskii82} Lyubarskii, Y.~E., \& Sunyaev, R.~A.\ 1982, Soviet Astronomy Letters, 8, 330

\bibitem[Maeda et al.(2007)]{Maeda07_nucleo_aj_nomoto} Maeda, K., et al.\ 2007, \apjl, 658, L5

\bibitem[Malesani et al.(2004)]{Malesani04} Malesani, D., et al.\
2004, \apjl, 609, L5

\bibitem[Malesani et al.(2009)]{Malesani09_08d_spectr} Malesani, D., et al.\ 2009, \apjl, 692, L84

\bibitem[Matzner
\& McKee(1999)]{Matzner99} Matzner, C.~D., \& McKee, C.~F.\ 1999, \apj, 510, 379

\bibitem[Mazzali et al.(2006)]{Mazzali06_prog_aj} Mazzali, P.~A., et al.\ 2006, \nat, 442, 1018

\bibitem[Mazzali et al.(2007)]{Mazzali07_no_pol_aj} Mazzali, P.~A., et al.\ 2007, \apj, 661, 892

\bibitem[Mazzali et al.(2008)]{Mazzali08_08D_SN} Mazzali, P.~A., et al.\ 2008, Science, 321, 1185

\bibitem[Modjaz et al.(2006)]{Modjaz06_aj_SN_analysis} Modjaz, M., et al.\ 2006, \apjl, 645, L21

\bibitem[Modjaz et al.(2009)]{Modjaz08_08D} Modjaz, M., et al.\
2009, \apj, 702, 226

\bibitem[Pian et al.(2006)]{Pian06_SN_aj} Pian, E., et al.\ 2006, \nat, 442, 1011

\bibitem[Sakurai
(1960)]{Sakurai60} Sakurai, A. \ 1960, Comm. Pure Appl. Math., 13,353

\bibitem[Soderberg et al.(2004)]{Soderberg04_bw_nojet} Soderberg, A.~M., Frail, D.~A., \& Wieringa, M.~H.\ 2004, \apjl, 607, L13

\bibitem[Soderberg et al.(2006)]{Soderberg06_rel_jet_aj} Soderberg, A.~M., et al.\ 2006, \nat, 442, 1014

\bibitem[Soderberg et al.(2008)]{Soderberg08} Soderberg, A.~M., et
al.\ 2008, \nat, 453, 469

\bibitem[Stanek et al.(2003)]{Stanek03} Stanek, K.~Z., et al.\
2003, \apjl, 591, L17

\bibitem[Stratta et
al.(2007)]{Stratta07} Stratta, G., et al.\ 2007, \aap, 461, 485

\bibitem[Svensson(1984)]{Svensson84} Svensson, R.\ 1984, \mnras,
209, 175

\bibitem[Tan et al.(2001)]{Tan01} Tan, J.~C., Matzner, C.~D., \& McKee, C.~F.\ 2001, \apj, 551, 946

\bibitem[Tanaka et al.(2008)]{Tanaka08_08D_prog_nomoto} Tanaka, M., et al.\ 2008, arXiv:0807.1674

\bibitem[Wang et al.(2007)]{Wang07} Wang, X.-Y., Li, Z.,
Waxman, E., \& M{\'e}sz{\'a}ros, P.\ 2007, \apj, 664, 1026

\bibitem[Waxman(2004a)]{Waxman_bw1} Waxman, E.\ 2004, \apj, 602, 886

\bibitem[Waxman(2004b)]{Waxman_bw2} Waxman, E.\ 2004, \apjl, 605, L97

\bibitem[Waxman \& Loeb(1999)]{WnL99} Waxman, E., \& Loeb, A.\ 1999, \apj, 515, 721

\bibitem[Waxman \& Loeb (2001)]{WL01} Waxman, E., \& Loeb, A.\ 2001, Physical Review Letters, 87, 071101

\bibitem[Waxman et al.(2007)]{Waxman07} Waxman, E.,
M{\'e}sz{\'a}ros, P., \& Campana, S.\ 2007, \apj, 667, 351

%\bibitem[Weaver
%\& Chapline(1974)]{1974ApJ...192L..57W} Weaver, T.~A., \& Chapline, G.~F.\ 1974, \apjl, 192, L57

\bibitem[Weaver(1976)]{Weaver76} Weaver, T.~A.\ 1976, \apjs, 32,
233

\bibitem[Zel'dovich \& Raizer(1966)]{Zel'dovich66} Zel'dovich, Ya. B. \& Raizer, Yu. P., Physics of Shock Waves and High-Temperature Hydrodynamic Phenomena, Dover Publications, Inc.




\end{thebibliography}

\end{document}